\documentclass[10pt,twocolumn,pre,byrevtex,superscriptaddress,showpacs,bibnotes]{revtex4}
\usepackage{graphics}

\begin{document}


\title
{Compromise and Synchronization \\in Opinion Dynamics}

\author
{Alessandro Pluchino$^{1}$, Vito Latora$^{1}$, Andrea Rapisarda$^{1}$}

\address
{$^{1}$Dipartimento di Fisica e Astronomia, Universit\`a di
Catania, \\ and INFN sezione di Catania, Via S. Sofia 64, 95123
Catania, Italy}

\begin{abstract}
We discuss two models of opinion dynamics. First we present a
brief review of the Hegselmann and Krause (HK) compromise model in
two dimensions, showing that it is possible to simulate the
dynamics in the limit of an infinite number of agents by solving
numerically a rate equation for a continuum distribution of
opinions. Then, we discuss the Opinion Changing Rate (OCR) model,
which allows to study under which conditions a group of agents
with a different natural tendency (rate) to change opinion can
find the agreement. In the context of the this model, consensus is
viewed as a synchronization process.
\end{abstract}


\maketitle 

Since  the behavioral revolution and the birth of cybernetics, the
so called '{\it soft}' social sciences have emulated both the
intellectual and methodological paradigms of the '{\it strong}'
natural sciences \cite{chaos_soc}. The certainty and stability of
the Newtonian paradigm has represented for decades the cornerstone
of sciences like psychology, economy and sociology, which have
been largely inspired by classical mechanics and statistical
thermodynamics. Clearly this trend has continued since quantum
mechanics, chaos and complexity revolutions have leaded to a
reconsideration of the relevance of the Newtonian paradigm to all
natural phenomena.
\\
In the last years, new disciplines such as econophysics and
sociophysics have largely demonstrated the power of {\it
agent-based} computational models in simulating complex adaptive
systems (financial markets, cultural evolution, social structures,
voter communities) in which large numbers of individuals are
involved in massively parallel local interactions
\cite{abs,stauff1}. In agent-based models, individuals are
modelled as autonomous interacting agents with a variable degree
of internal complexity and  numerical simulations represent
computational experiments to study the evolution of a given social
system under controlled  conditions. Of course in many cases the
individual cognitive behavior is oversimplified, as for example in
opinion dynamics models where human opinions are reduced to
integer or real numbers \cite{HK,SDG}. In more complicated models,
individuals are simulated by means of simple neural networks or
associative memories. Although  in this case, many of the
simplifications adopted are somehow unrealistic. On the other
hand, also the Kepler's laws assumption of Earth as a point-mass
was not realistic at all, but for the purpose of describing
celestial motion it turned out very successful. Furthermore, the
aim of agent-based simulations is to provide information on
averages over many people, and not on the fate of a specific
person. In this sense, despite of their simplicity, these models
seem to work very well. For example, the Sznajd model prediction
of the distribution of votes among candidates in Brazilian and
Indian elections is encouraging, although the model is not able to
predict the number of votes one specific candidate gets in one
specific election \cite{stauff2}.
\\
In the first part of this paper we discuss one of the well known
models of opinion dynamics, the so called {\it compromise model}
of Hegselmann and Krause \cite{HK}, showing some recent results
about its continuum version in a two-dimensional opinion space
\cite{vector}. In the second part of the paper we discuss a new
perspective in opinion dynamics based on the suggestion of a
possible role of synchronization in opinion formation. By means of
the so called {\it Opinion Changing Rate model}, a modified
version of the Kuramoto model adapted to a social context, we
study under which conditions a group of agents with a different
natural tendency ($rate$) to change opinion can find agreement
\cite{ocr}.

\section{Discrete and Continuum Opinion Dynamics in the 2-vector HK Consensus Model}

For the sociologist Robert Axelrod ``culture'' is modelled as an
array of features, each feature being specified by ``traits'',
which are expressed by numbers. The number of features or
dimensions is nothing but the number of components of a vector,
and two persons interact if and only if they share at least one
common feature (i.e. the same value of the corresponding vector
component) \cite{Axel}. In this model, two persons are culturally
closer the more features they have in common, and the number of
these common features is, in turn, related to the probability for
the two individuals to interact. Starting from the Axelrod model,
several simple agent-based models of opinion formation have been
devised, mostly by physicists \cite{stauff1,HK,SDG}.
\\
In general, a typical scalar opinion formation model starts by
assigning randomly a real number (chosen in a given interval) to
every agent of the system. Then the dynamics starts to act, and
the agents rearrange their opinion variables, due to their
interactions. At some stage, possibly, the system reaches a
configuration which is stable under the dynamics. This final
configuration may represent consensus, when all agents share the
same opinion, polarization, when there are two main clusters of
opinions ("parties"), or fragmentation, when several opinion
clusters survive. However, a discussion between two persons is not
simply stimulated by their common view/preference about a specific
issue, but it in general depends on the global affinity of the two
persons, which is influenced by several factors. So, for a more
realistic modelling of opinion dynamics, one should represent the
opinions/attitudes like vectors (as in the Axelrod model), and not
like scalars. In this section we will focus on the 2-vector
version of the Hegselmann and Krause (HK) compromise model and we
will show that it is possible to simulate the discrete opinion
dynamics in the limit of an infinite number of agents by solving
numerically a rate equation for a continuum distribution of
opinions \cite{vector}.
\\
The HK model \cite{HK} is based on the concept of bounded
confidence, i.e. on the presence of a parameter $\epsilon$, called
{\it confidence bound}, which expresses the compatibility among
the agents in the {\it opinion space}. If the opinions of two
agents $i$ and $j$ differ by less than $\epsilon$, their positions
are close enough to allow for a discussion, which eventually leads
to a change in their opinions, otherwise the two agents do not
interact with each other. The {\it physical space} occupied by the
agents living in a society or a community can be modelled as a
graph, where the vertices represent the agents and the edges
relationships between agents. So we say that two agents can
eventually talk to each other if there is an edge joining the two
corresponding vertices (in graph language, if the two vertices are
neighbors). In the following we will consider only the general
case of a society where everybody talks to everybody.
%
%
\begin{figure}  [t]
\begin{center}
\resizebox{3.5in}{!}{\includegraphics{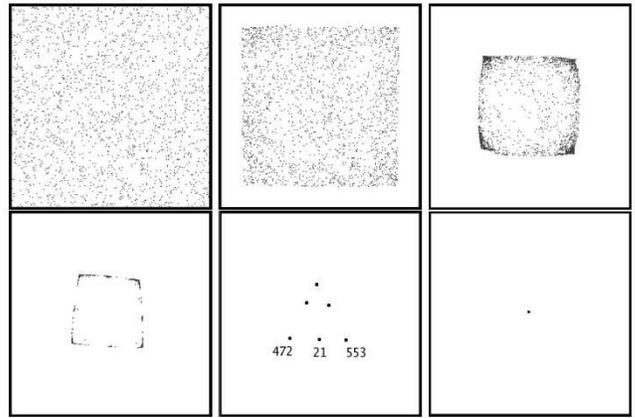}}
\caption{Sequence of snapshots of the 2D squared opinion space for
the 2-vector discrete HK model with $N=2000$ agents and
$\epsilon=0.25$. The points in the upper-left panel represent
different randomly distributed opinions at t=0. In the other
panels, where we show successive (but not consecutive) time steps
of a Monte Carlo simulation with simultaneous update, opinions
merge together in different clusters. Finally, in the lower-left
panel (t=12) the consensus is reached and all the opinions occupy
the same position. In the second-last panel, the number of
opinions concentrated in the bottom clusters is also indicated
(see text).}
\end{center}
\end{figure}
The dynamics of the HK model is usually simulated by means of
Monte Carlo (MC) algorithms. One chooses at random one of the
agents and checks how many of its neighbors (in the physical
space) are compatible, i.e. lie inside the confidence range in the
opinion space. Next, the agent takes the average opinion of its
compatible neighbors. The procedure is repeated by selecting at
random another agent and so on. The type of final configuration
reached by the system depends on the value of the confidence bound
$\epsilon$. For a scalar opinion space $[0,1]$ it has been shown
that consensus is reached for $\epsilon>\epsilon_c$, where the
critical threshold $\epsilon_c$ is strictly related to the type of
graph adopted to model society: actually, it can take only one of
two possible values, $\epsilon_c\sim 0.2$ and $0.5$, depending on
whether the average degree of the graph (i.e. the average number
of neighbors) diverges, as in our case of a completely connected
graph, or stays finite when the number of vertices goes to
infinity \cite{san1}. On the other hand, the 2-vector HK model on
a completely connected graph is much less studied than the
1-dimensional version. In this case the opinion space is
represented by the points $(x,y)$ of a bidimensional manifold,
that in general is a square $[0,1]\times [0,1]$ and the confidence
range is a circle whose radius is the confidence bound $\epsilon$.
\begin{figure}[t]
\begin{center}
\resizebox{3.5in}{!}{\includegraphics{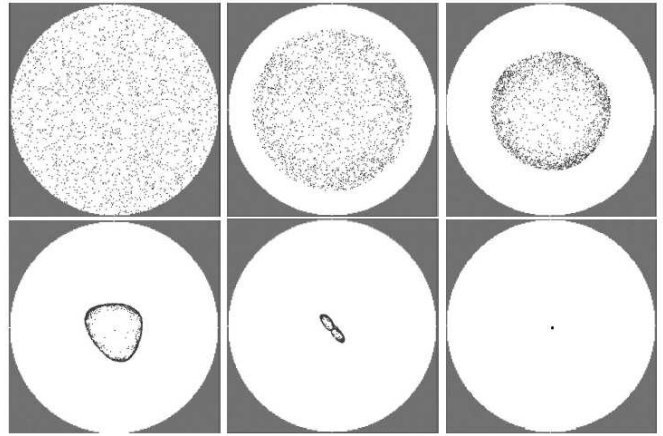}}
\caption{Sequence of snapshots of the 2D circular opinion space
for the 2-vector discrete HK model with $N=2000$ agents and
$\epsilon=0.25$. The dynamics is the same than in Fig.1. The
different shape of the space influences the dynamical evolution of
the opinions and consensus is reached in a shorter time (t=6) than
in the case of Fig.1. }
\end{center}
\end{figure}
\\
In Fig.1 we plot a sequence of snapshots of the opinion space for
a discrete configuration of $N=2000$ agents and a value $\epsilon
=0.25$. Each point in the first snapshot (upper-left panel)
represents the opinion of one agent at time t=0. Then the system
evolves by means of simultaneous updates (i.e. all the opinions
are updated at each MC time step) merging the opinions in bigger
and bigger clusters until a stationary state is reached. After 12
time steps (lower-right panel) consensus is fully obtained and all
the opinions lie on the same cluster. More in general, from
extended numerical simulations it results \cite{vector} that the
consensus threshold for the discrete 2-vector HK model is
$\epsilon_c\sim 0.24$, a value slightly greater than that one
found for the scalar model with the same topology. This value
tends to the value $\epsilon_c\sim 0.23$ when the number of agents
grows.
\\
If we look at the basis of the triangle in the second-last
snapshot of Fig.1, just before reaching final consensus, we can
see that the two big clusters at the vertices, made by around
$500$ opinions and  at a reciprocal distance greater than the
confidence bound (thus {\it a-priori} not interacting), are  in
contact only by means of a small cluster of $21$ agents. Such a
phenomenon is very frequent in the Monte Carlo simulations of the
HK model when the system is approaching consensus (in both one and
two dimension); indeed, almost always consensus is reached only
because of this phenomenon. This models an important feature of
real social networks, i.e. the existence of the so called {\it
connectors} which play the role of a bridge between otherwise not
interacting social groups, thus ensuring the cohesiveness of the
entire network \cite{gladwell}.
\\
Another peculiar feature of the HK model, clearly visible in the
upper snapshots of Fig.1, is the fact that the dynamics always
starts to act from the edges of the opinion space, where the
opinion distribution is necessarily inhomogeneous, so that it is
essentially the shape of the opinion space which rules the
symmetry of the resulting cluster distribution. In order to better
appreciate this effect, we plot in Fig.2 the temporal evolution of
the same system of Fig.1, but with a circular opinion space. In
this case, even if the final configuration is the same as before,
the resulting dynamics is different and, for the same value of the
confidence bound, consensus is reached more quickly (six MC time
steps with simultaneous update) due to the greater symmetry of the
opinion space. The circular symmetry has a remarkable effect also
on the consensus threshold, that in this case tends to that of the
corresponding scalar HK model, i.e. $\epsilon\sim 0.2$.
\begin{figure}[t]
\begin{center}
\resizebox{3.5in}{!}{\includegraphics{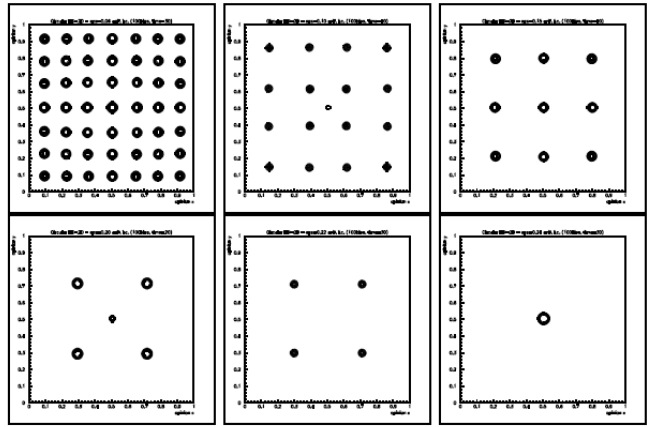}}
\caption{Final configurations of the 2-vector HK model with a
continuum distribution of opinions. From top left to bottom right:
$\epsilon$=0.08,0.10,0.15; 0.20,0.22,0.24. }
\end{center}
\end{figure}
In a recent paper \cite{vector} it has been shown that the
2-vector HK model on a completely connected graph and with a
squared opinion space can be described by means of a rate equation
for a continuum distribution of opinions $P(x,y,t)$. The rate
equation can be solved numerically, as already done for the scalar
compromise model of Deffuant et al. \cite{bennaim}). The
advantages of the evolution equation over discrete Monte Carlo
simulations are that one can directly deal with a system with an
arbitrarily large number of agents, and that the final cluster
configurations for a continuum distribution are much more
symmetric and regular, thus allowing a better resolution of the
progressive merging of opinion clusters.
\\
This is clearly visible in Fig.3, where we show the final
configurations of the opinion space for several values of the
confidence bound. In this case the squared ($x,y$) opinion space
has been reduced to a grid of $100 \times 100$ bins. All the
simulations start from a flat distribution $P(x,y,t=0)={\rm
const}$ and the dynamics runs until the distribution $P(x,y,t)$
reaches a stationary state for a given value of the confidence
bound \cite{vector}.
\\
As one can see that, for small value of $\epsilon$, a regular
lattice of clusters appears, with a squared shape inherited by the
shape of the opinion space  (as happened for the discrete HK
model). Going on, for greater values of $\epsilon$, one can
observe the progressive merging of the pairs of clusters with
 reciprocal distance less than the confidence bound radius. Finally, above the
critical threshold $\epsilon_c\sim 0.23$, consensus is completely reached.
This result confirms the threshold value found with the MC simulations
for a discrete dynamics of opinions in the limit of a large number of agents
and encourages further applications of the rate equation technique to
other opinion formation models.

\begin{figure}[t]
\begin{center}
\resizebox{3.5in}{!}{\includegraphics{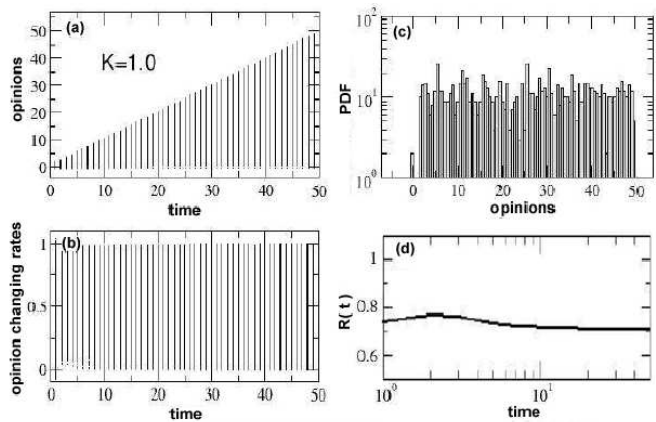}}
\caption{OCR model for $N=1000$ and $K=1.0$ (incoherent phase).
In panels (a) and (b) we report, for each time step, 1000
points corresponding respectively to
the opinions and opinion changing rates of the N agents,
while in panels (c) and (d) we show, respectively,
the final distribution of opinions and the
time evolution of the order parameter $R$.
}
\end{center}
\end{figure}

\section{The Opinion Changing Rate model: a role for synchronization in opinion formation}

Most of the opinion formation models, as for example  the HK model
presented in the previous section, have the limitation of not
taking into account the individual inclination to change, a
peculiar feature of any social system. In fact, each one of us
changes ideas, habits, style of life or way of thinking in a
different way, with a different velocity. There are conservative
people that strongly tend to maintain their opinion or their style
of life against everything and everyone. There are more flexible
people that change ideas very easily and follow the current
fashions  and trends. Finally,  there are those who run faster
than the rest of the world anticipating the others. These
different tendencies can be interpreted as a continuous spectrum
of different degrees of natural inclination to changes.
\\
In a recent paper \cite{ocr} we have showed how such an individual
inclination to change, differently distributed in a group of
people, can affect the opinion dynamics of the group itself. If we
switch from  the question: ``Could agents with initial different
opinions reach a final  agreement ?'' into the more realistic  one
`` Could agents with a  different
 natural tendency to change opinion
 reach a final agreement ? '', we can introduce
a new concept, the natural
opinion changing rate, that is very similar to the
characteristic frequency of an oscillator. In such a way,
we can treat consensus as a peculiar kind of synchronization
(frequency locking), a phenomenon which has been very
well studied in different contexts
by means of the Kuramoto model\cite{kuramoto_model}.
\\
The Kuramoto model of coupled oscillators is one of the simplest and most
successful models for synchronization.
It is simple enough to be analytically solvable, still retaining
the basic principles to produce a rich variety of dynamical regimes
and synchronization patterns.
The dynamics of the model is given by
\begin{equation}
    \dot{\theta_i} (t)  = \omega_i + \frac{K}{N} \sum_{j=1}^N
      \sin ( \theta_j  - \theta_i )  ~~~~~i=1,\dots,N
\label{kuramoto_eq1}
\end{equation}
where $\theta_i (t)$ is the phase (angle) of the $i$th oscillator
at time $t$ ($-\pi<\theta_i(t)<\pi$), while $\omega_i$ is its
intrinsic frequency randomly drawn from a symmetric, unimodal
distribution $g(\omega)$ with a first moment $\omega_0$ (typically
a Gaussian distribution or a uniform one). These natural
frequencies $\omega_i$  are time-independent. The sum in the above
equation is running over all the oscillators so that this is an
example of a globally coupled system. The parameter $K \geq 0$
measures the coupling strength in the global coupling term. For
small  values of $K$, each oscillator tends to run independently
with its own frequency,  while for large values of  $K$, the
coupling tends to synchronize (in phase and frequency) the
oscillator with all the others. Kuramoto showed that the model,
despite the difference in the natural frequencies of the
oscillators, exhibits a spontaneous transition from incoherence to
collective synchronization, as the coupling strength is increased
beyond  a certain threshold $K_c$ \cite{strogatz}.
\begin{figure}[t]
\begin{center}
\resizebox{3.5in}{!}{\includegraphics{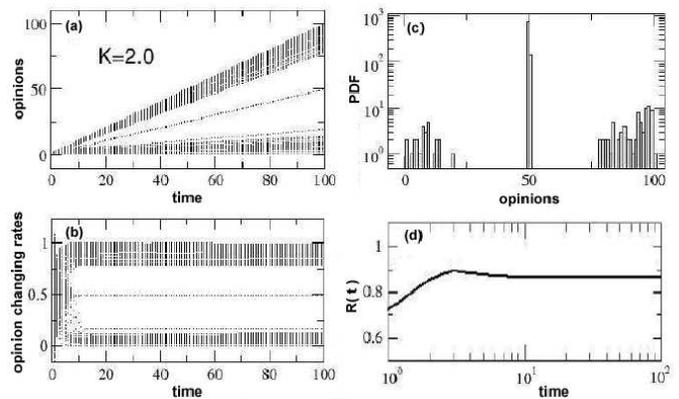}}
\caption{OCR model for $N=1000$ and $K=2.0$ (partially coherent phase).
Same quantities as in Fig.4.
}
\end{center}
\end{figure}
\\
The existence of a critical threshold for synchronization in
Kuramoto model is very similar to the consensus threshold found in
the majority of the opinion formation models. Of course, at
variance with the phases in the Kuramoto model, in a model for
opinion dynamics we do not need periodic opinions nor limited
ones: in fact, the opinions have a very general meaning and can
represent the style of life, the way of thinking or of dressing
etc. Thus we do not consider periodic boundary conditions and we
assume  $x_i \in ] -\infty , +\infty[ ~~ \forall i=1,...,N$. The
dynamics of the OCR model is governed by the following set of
differential equations \cite{ocr}:
\begin{equation}
    \dot{x_i} (t)  = \omega_i + \frac{K}{N} \sum_{j=1}^N\alpha
      \sin ( x_j  - x_i ) e^{- \alpha |x_j  - x_i| }  ~~~~~i=1,\dots,N
\label{OCR_eq1}
\end{equation}
Here $x_i (t)$ is the opinion (an unlimited real number) of the
$i$th individual at time $t$, while $\omega_i$ represents the so
called \textit{natural opinion changing rate}, i.e. the intrinsic
inclination, or natural tendency, of each individual to change his
opinion (corresponding to the {\it natural frequency} of each
oscillator in the Kuramoto model). As in the Kuramoto model, also
in the OCR model the $\omega$'s are randomly drawn from a given
symmetric, unimodal distribution $g(\omega)$ with a first moment
$\omega_0$. Usually a uniform distribution centered at $\omega_0$
is  used. In this way one simulates the fact that in a population
there are: 1) conservative individuals, that naturally tend to
change their opinion very slowly, and thus are characterized by a
value of $\omega_i$ smaller than $\omega_0$; 2)  more flexible
people, with $\omega_i \sim \omega_0$, that change idea more
easily and follow the new fashions and trends; 3) individuals with
a value of $\omega_i$ higher than $\omega_0$, that run faster than
the others in suggesting new ideas and insights.
\\
In the equation (\ref{OCR_eq1}) $K$, as usual, is the coupling
strength. The exponential factor in the coupling term ensures
that, for reciprocal distance higher than a certain threshold,
tuned by the parameter $\alpha$, opinions will no more influence
each other. Such a requirement is inspired by the confidence bound
concept discussed in the previous section. (Please note that, due to a misprinting, there is an $\alpha$
factor missing in the coupling term of Eq.(7) of ref.\cite{ocr}).   At this point we can
study the opinion dynamics of the OCR model by solving numerically
the set of ordinary differential equations (\ref{OCR_eq1}) for a
given distribution of the $\omega$'s (natural opinion changing
rates) and for a given coupling strength $K$. In particular, we
want to find out if, as a function of $K$, there is a transition
from an incoherent phase, in which people change opinion each one
with his natural rate $\omega_i$, to a synchronized one in which
all the people change opinion with the same rate and share a
common social trend, a sort of 'public opinion'.
\begin{figure}[t]
\begin{center}
\resizebox{3.5in}{!}{\includegraphics{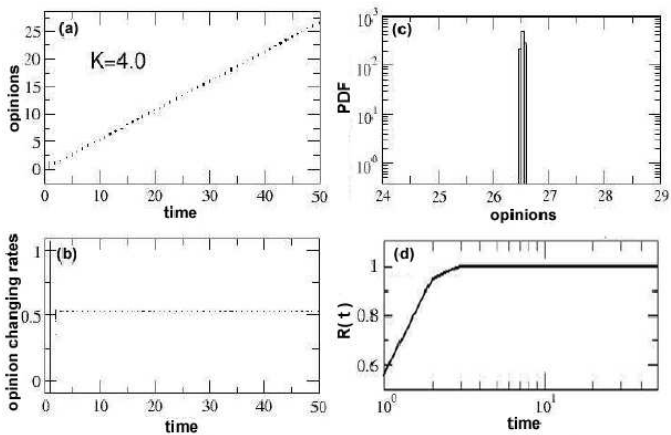}}
\caption{OCR model for $N=1000$ and $K=4.0$ (synchronized phase).
Same quantities as in the previous figures.
}
\end{center}
\end{figure}
In order to measure the degree of synchronization of the system we
decided to adopt an order parameter $R(t)$ related to the standard
deviation of the opinion changing rate $\dot{x_j}(t)$ and defined
as $R(t) = 1 - \sqrt{ \frac{1}{N} \sum_{j=1}^N (\dot{x}_j(t) -
\dot{X}(t))^{2}}$, where $\dot{X}(t)$ is the average over all
individuals of $\dot{x_j}(t)$. It is easy to see that $R=1$ in the
fully synchronized phase, where all the agents have exactly the
same opinion changing rate (and very similar opinions), while
$R<1$ in the incoherent or partially synchronized phase, in which
the agents have different opinion changing rates and different
opinions. The numerical simulations have been performed typically
with N=1000 agents and with an uniform distribution of the initial
individual opinions $x_i(t=0)$ in the range [-1,1]. The natural
opinion changing rates $\omega_i$ are taken from a uniform
distribution in the range [0,1]. We fix the value of the coupling
$K$ and we let the system to evolve until a stationary
(asymptotic) value $R_\infty$ for the order parameter is obtained.
In this way it is easy to recognize a Kuramoto-like transition
from an incoherent phase (for $K<K_c\sim1.4$) to a partially
coherent (for $K\in[1.4,4.0]$) and, finally, to a fully
synchronized phase (for $K>4.0$) \cite{ocr}.
We now focus on the details of the dynamical evolution in each of
the three phases.
\\
In Fig.4 we show the case of very small coupling, $K=1.0$. In the
left part we show the time evolution of the opinions and of the
opinion changing rates (angular velocities or frequencies). In the
right part, instead, we plot the final distribution of opinions
and the order parameter time evolution. Because of the weak
interactions we are in the incoherent phase and each agent tends
to keep his natural opinion changing rate. It follows that the
different opinions diverge in time without reaching any kind of
consensus. In correspondence, the order parameter R takes the
minimum possible value that, at variance with the Kuramoto model,
is not zero. We could look at this case as to an $'anarchical'$
society.
\\
In Fig.5 we plot the same quantities than before but in the case
$K=2.0$. The coupling is still weak but strong enough to give rise
to three different clusters of evolving opinions, each with a
characteristic changing rate: the largest number of the agents,
representing what we could call the "public opinion", moves with
an intermediate rate along the opinion axis, but there is a
consistent group of people remaining behind them and also a group
of innovative people (quicker in supply new ideas and ingenuity).
From a political point of view, we could interpret this situation
as a sort of 'bipolarism' with a large number of 'centrists'. In
this case the order parameter is larger than in the previous
example, but still less than one since the opinion synchronization
is only partial.
\begin{figure}[t]
\begin{center}
\resizebox{3.5in}{!}{\includegraphics{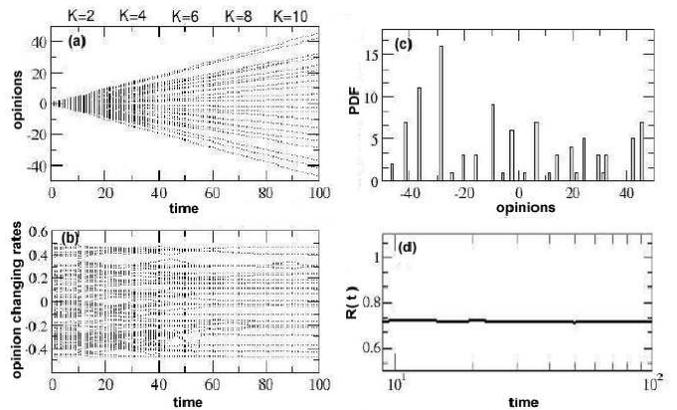}}
\caption{
OCR model for $N=100$ and a value of
$K$ increasing with a constant rate from 0.1
up to 10.1. Same quantities as in the previous figures.
}
\end{center}
\end{figure}
\\
Finally, in Fig.6 we report the case $K=4.0$.  Here the coupling
is so strong that all the opinions change at the same rate and we
observe a single final cluster in the opinion distribution. In
this $'dictatorial'$ society all the agents think in the same way
and follow the same trends and fashions. Although the $natural$
frequencies of the agents are - as in the previous examples -
different from each others, their opinion changing rates rapidly
synchronize (frequency locking) and thus the order parameter R
reaches a saturation value equal to one.
\\
Summarizing, it has been found \cite{ocr} that in order to ensure
a 'bipolarism' - i.e. an equilibrium between conservative and
progressist components - a changing society needs a level of
coupling $K$ strictly included in a narrow window ($1.5 < K <
2.5$) inside the partially synchronized phase. Otherwise such an
equilibrium will be broken and the final result will be anarchy or
dictatorial regime. But it is worth to observe that these
conclusions have been obtained for systems with fixed coupling
$K$, simulating societies with a stable degree of interconnections
among their members. Thus it is interesting to explore what
happens if the coupling is let to increase its value during the
dynamics, in order to simulate a society in which the
interconnections between the agents increase in time, due for
example to the improvement in transport or in communications.
\\
In Fig.7 we show the results for a system in which the coupling is
uniformly increased from $K=0.1$ to $K=10.1$. The agents' opinions
initially spread freely, and then rapidly freeze in a large number
of non-interacting clusters with different changing rates and
variable sizes. Actually, it results that this particular cluster
distribution, that could be socially interpreted as a
multipolarism, cannot be obtained in simulations with a constant
coupling. This could suggest that the increase of interactions
between the members of a society is crucial to stabilize a
plurality of different non-interacting clusters of opinions
(different ideologies, political parties, etc.) typical of a
multipolar democracy. It seems to suggest also that a stable
bipolarism is possible only in societies with a fixed degree of
internal interconnections.

\section{Conclusions}
In this paper we have shown that even simple opinion formation
models are able to capture many general features of real social
systems. In the first part we have discussed an extension of the
scalar opinion dynamics of Hegselmann-Krause model to the case in which
the opinion is not just a scalar but a 2-vector with real-valued
components. We investigated a community where everybody talks to
everybody by means of Monte Carlo simulations of discrete opinion
distributions (for squared or circular opinion space) and by
solving numerically a rate equation for a continuum opinion
distribution. By studying the consensus thresholds we found that
the continuum case can be considered as the limit of the discrete
one for a great number of agents. In the second part of the paper
we have discussed a social variation of the Kuramoto model, the so
called Opinion Changing Rate model (OCR). The concept of
\textit{'opinion changing rate'} transforms the usual approach to
opinion consensus into a synchronization problem. As for the
Kuramoto model, the OCR model exhibits a phase transition from an
incoherent phase to a synchronized one and shows many interesting
features with a clear social meaning.


\end{document}